# Detecting the rapidly expanding outer shell of the Crab Nebula: where to look

Xiang Wang[1], G. J. Ferland[1], J. A. Baldwin[2], E. D. Loh[2], C. T. Richardson[2]

[1]Department of Physics and Astronomy, University of Kentucky, Lexington, KY 40506, USA

xiang.wang@uky.edu

[2]Department of Physics and Astronomy, Michigan State University, East Lansing, MI 48824-2320, USA

# Abstract

We present a range of steady-state photoionization simulations, corresponding to different assumed shell geometries and compositions, of the unseen postulated rapidly expanding outer shell to the Crab Nebula. The properties of the shell are constrained by the mass that must lie within it, and by limits to the intensities of hydrogen recombination lines. In all cases the photoionization models predict very strong emission from high ionization lines that will not be emitted by the Crab's filaments, alleviating problems with detecting these lines in the presence of light scattered from brighter parts of the Crab. The NIR [Ne VI] λ7.652 μm line is a particularly good case; it should be dramatically brighter than the optical lines commonly used in searches. The C IV λ1549Å doublet is predicted to be the strongest absorption line from the shell, which is in agreement with HST observations. We show that the cooling timescale for the outer shell is much longer than the age of the Crab, due to the low density. This means that the temperature of the shell will actually "remember" its initial conditions. However, the recombination time is much shorter than the age of the Crab, so the predicted level of ionization should approximate the real ionization. In any case, it is clear that IR observations present the best opportunity to detect the outer shell and so guide future models that will constrain early events in the original explosion.

Subject headings: ISM: supernova remnants – star: supernovae: individual: SN1054 – methods: numerical

# 1. Introduction

The Crab Nebula is generally thought to have been produced by a core collapse supernova. The total mass in the observed ejecta is 2-5 $M_{sun}$ (Davidson & Fesen 1985; Fesen, Shull & Hurford 1997) and the pulsar should have a mass of about $1.4 M_{sun}$ (Davidson & Fesen 1985). This is much less than the total mass of 8-13 $M_{sun}$ (Nomoto 1985, 1987; Kitaura, Janka & Hillebrandt 2006) thought to be in the star before the explosion. Thus the long-standing problem, where is the missing mass? The possibility most often discussed is that it lies within an unseen outer shell, sometimes referred to as the Crab's halo. The literature on this is comprehensive, with Lunqdvist & Tziamtzis (2012) and Smith (2013) giving good summaries of the current situation. Smith (2013) discusses an alternative explanation, that the Crab was a type of under-luminous supernova.

There have been only a few predictions of the detailed spectrum of the outer shell. Lundqvist, Fransson & Chevalier (1986) did time-dependent numerical simulations of the spectrum with a constant density structure and Sankrit &

Hester (1997) predict some properties of a photoionized and shock heated shell. Here we use an up-to-date atomic database in the spectral synthesis code Cloudy (Ferland et al. 2013) to compute emission and absorption spectra. We largely confirm previous estimates of the hydrogen emission but find that strong optical and infrared coronal lines should also be present. We identify promising lines in the IR that would be a robust indicator of the presence of this outer shell.

## 2. Parameters of the outer shell

The total luminosity of the Crab Nebula, and its spectral energy distribution (SED), are well known (Davidson & Fesen 1985). Although other energy sources such as shocks may be present (Sankrit & Hester 1997), photoionization by this continuum must occur (the SED is observed) and by itself can power the outer shell. Shock heating would only add to this. To compute a photoionization model of the outer shell and its spectrum we must specify the gas composition, its density, and how the density varies with radius.

We assume that the outer shell is an inhomogeneous shell with an uncertain outer radius, but with an inner radius equal to the outer radius of the familiar Crab, $R_{in}$= 5.0×10$^{18}$ cm (Sankrit & Hester 1997). The expansion velocity at the inner radius $v_{in}$ is roughly 1680 km s$^{-1}$ at this radius (the Crab is, of course, not a sphere, so this is a simplification).

A velocity gradient must be present, since the outer shell lies outside the familiar Crab. We consider both a Hubble flow, with $v(r) \propto r$, and an arbitrary velocity law as a sensitivity test, with $v(r) \propto r^2$. We obtain two different density laws from these two velocity distributions and apply them in this paper to check how predictions depend on this assumption.

The total mass in the outer shell may be of order 4 to 8 $M_{sun}$ (Sollerman et al. 2000). We assume $4 M_{sun}$ recommended by Sollerman et al. (2000), which we show below is consistent with limits to the line surface brightness (Fesen, Shull & Hurford 1997; Tziamtzis et al. 2009). We combine this with the three power laws given above to find the gas density as a function of radius.

### 2.1. The outer radius

We will determine the gas density by combining the total mass with the density law and the inner and outer radii. The outer radius is unknown but must be specified to determine the gas density. Given our assumptions about the radius – velocity law, the outer radius corresponds to a particular highest expansion velocity. Chevalier (1977) gives a range of expansion velocities between 5,000 km s$^{-1}$ and 10,000 km s$^{-1}$, Lundqvist, Fransson, & Chevalier (1986) give a maximum expansion velocity of 5,000 km s$^{-1}$, Sankrit & Hester (1997) assume a maximum velocity of 10,000 km s$^{-1}$, and Sollerman et al. (2000) quote 6370 km s$^{-1}$. We assume the velocity at the outer radius $v_{out}$ = 6370 km s$^{-1}$, a velocity ~3.8 times larger than the expansion of the observed nebula, and give results relative to this velocity. The $v(r) \propto r$ Hubble flow results in

$$R_{out} \approx 3.8 R_{in} = 1.9 \times 10^{19} [\text{cm}]. \tag{1}$$

while for the $v(r) \propto r^2$ expansion law the outer radius is

$$R_{out} \approx 1.9 R_{in} = 9.5 \times 10^{18} [\text{cm}]. \tag{2}$$

## 2.2. The density law

For the spectroscopic simulations we need to set the outer shell density $n_0$ at its inner edge $R_{in}$, the density law $n(r) \propto r^\alpha$, and the outer radius $R_{out}$. We investigate two density laws here, $\alpha = -3$ and $\alpha = -4$. The density law is determined by two quantities, how the expansion velocity varies with radius, $v(r) \propto r^\gamma$, and how the mass flux varies with radius, $MF \propto r^\beta$. We consider three cases, summarized in Table 1, as follows:

(I) The simplest case is a Hubble-law expansion, the sudden release of mass with a range of velocities so that $\gamma = 1$ and $v \propto r$. For the mass flux, the simplest assumption is that the initial density distribution is constant, so that

$$MF = 4\pi r^2 n(r) v(r) \propto 4\pi r^{2+\alpha+\gamma} = 4\pi r^\beta \tag{3}$$

is constant. Since $\gamma = 1$, if $\alpha = -3$, then $\beta = 0$, indicating mass flux conservation.

(II) As the second case we still assume that the Hubble velocity law is maintained so that $\gamma = 1$ and $v \propto r$. If $\alpha = -4$, then $\beta = -1$, meaning that the mass flux decreases with increasing radius. This may happen if the outer layer of the star had a lower density.

(III) As a third case we also consider $\alpha = -4$. As a sensitivity test, we will also test an arbitrarily different velocity law expansion with $\gamma = 2$ and $v(r) \propto r^2$. In this case we also obtain $\beta = 0$, that is, the mass flux is conserved.

The density law for case I is

$$n(r) = n_0 \left(\frac{r}{R_{in}}\right)^\alpha = n_0 \left(\frac{r}{R_{in}}\right)^{-3} [\text{cm}^{-3}] \tag{4}$$

and for case II and case III is

$$n(r) = n_0 \left(\frac{r}{R_{in}}\right)^{-4} [\text{cm}^{-3}]. \tag{5}$$

## 2.3. The shell mass and inner density

We can calculate $n_0$ by mass conservation,

$$M_{halo} = 4\pi \int_{R_{in}}^{R_{out}} m\, n_0 \left(\frac{r}{R_{in}}\right)^\alpha r^2 dr \; [\text{gm}]. \tag{6}$$

Here $M_{halo}$ is the total mass of the outer shell and $m$ is the mass per hydrogen for the assumed composition. Note that the composition of a supernova remnant is usually different in different parts, therefore we assume three different compositions for the outer shell: the abundances of some of the Crab filaments (Pequignot & Dennefield 1983), solar abundances (recommended by Sollerman et al. 2000), and ISM abundances (which are basically solar with grains). If $\mu$ is the mass of the proton then $m = 3.8\mu$ for the enhanced Crab abundances derived by Pequignot & Dennefield (1983) and $m = 1.4\mu$ for solar and ISM abundances. A list of assumed abundances is given in table 3a in Pequignot & Dennefield (1983). We obtain the following expression for $n_0$ with the middle value $m = 2.6\mu$ and $M_{halo} = 4\, M_{sun}$

$$\left. \begin{aligned} n_0 &= \frac{M_{halo}}{4\pi m R_{in}^3 \ln\frac{R_{out}}{R_{in}}} \\ &= 0.87 \frac{2.6\mu}{m} \frac{M_{halo}}{4 M_{sun}} \frac{\ln 3.8}{\ln\frac{R_{out}}{R_{in}}} [\text{cm}^{-3}] \end{aligned} \right\} \text{case I;} \tag{7}$$

$$\left. \begin{aligned} n_0 &= \frac{M_{halo}}{4\pi m R_{in}^3 \left(1 - \frac{R_{in}}{R_{out}}\right)} \\ &= 1.58 \frac{2.6\mu}{m} \frac{M_{halo}}{4 M_{sun}} \frac{0.74}{1 - \frac{R_{in}}{R_{out}}} [\text{cm}^{-3}] \end{aligned} \right\} \text{case II;} \tag{8}$$

$$\left. \begin{aligned} n_0 &= \frac{M_{halo}}{4\pi m R_{in}^3 \left(1 - \frac{R_{in}}{R_{out}}\right)} \\ &= 2.46 \frac{2.6\mu}{m} \frac{M_{halo}}{4 M_{sun}} \frac{0.47}{1 - \frac{R_{in}}{R_{out}}} [\text{cm}^{-3}] \end{aligned} \right\} \text{case III.} \tag{9}$$

We see that the density depends on both the inner radius and the outer radius for the $\alpha = -3$ law. This is important because the density determines the emission measure of the lines, and this depends on the uncertain outer radius. For the case of $\alpha = -4$, the density depends only on the inner radius if the outer radius is much larger than the inner radius. Table 5 in Sollerman et al. (2000) also gave the densities in the inner edge for different density laws.

### 2.4. Kinetic energy

Before proceeding with the model we derive the kinetic energy for each of these hypotheses. The kinetic energy of the filaments is about $3 \times 10^{49}$ ergs (Hester 2008), which is far less than the canonical $10^{51}$ ergs seen in the ejecta of core collapse supernovae (Davidson & Fesen 1985). We calculate the kinetic energy in the outer shell to check if this makes up the missing energy. We obtain the kinetic energy of the outer shell

$$E_k = 6.86 \times 10^{50} \frac{M_{halo}}{4M_{sun}} \frac{\ln 3.8}{\ln \frac{R_{out}}{R_{in}}} \frac{\left(\frac{R_{out}}{R_{in}}\right)^2 - 1}{13.44} [\text{erg}], \quad \text{case I};\tag{10}$$

$$E_k = 4.26 \times 10^{50} \frac{M_{halo}}{4M_{sun}} \frac{0.74}{1 - \frac{R_{in}}{R_{out}}} \frac{\frac{R_{out}}{R_{in}} - 1}{2.8} [\text{erg}], \quad \text{case II};\tag{11}$$

$$E_k = 6.70 \times 10^{50} \frac{M_{halo}}{4M_{sun}} \frac{0.47}{1 - \frac{R_{in}}{R_{out}}} \frac{\left(\frac{R_{out}}{R_{in}}\right)^3 - 1}{6.38} [\text{erg}], \quad \text{case III}.\tag{12}$$

These provide about half of the missing energy, which is within the uncertainty in our assumed shell parameters. Table 5 in Sollerman et al. (2000) also gave the kinetic energies of the outer shell for different density laws.

The next step is to predict the full emission and absorption line spectra of the outer shell using photoionization models.

### 2.5. Emission measure and line luminosity

We obtain the luminosities of emission lines from the numerical calculations presented below. We use H I line emissivities given by Osterbrock & Ferland (2006) (hereafter AGN3) and Ferland (1980). The luminosity of H$\beta$ is

$$L(\text{H}\beta) = \int \frac{4\pi j_{\text{H}\beta}}{n_e n_p} n(r)^2 dV \tag{13}$$

$$\approx \frac{4\pi j_{\text{H}\beta}}{n_e n_p} \times EM \ [\text{erg s}^{-1}] \tag{14}$$

where $\frac{4\pi j}{n_e n_p}$ is the H I Case B recombination coefficient (AGN3). EM is the volume emission measure, defined as

$$EM = \int n(r)^2 dV \tag{15}$$

$$\approx \int \left[n_0 \left(\frac{r}{R_{in}}\right)^\alpha\right]^2 dV \quad [\text{cm}^{-3}] \tag{16}$$

corresponding to

$$\left.\begin{aligned} EM &= \frac{4}{3}\pi n_0^2 R_{in}^3 \left[1-\left(\frac{R_{in}}{R_{out}}\right)^3\right] \\ &= 3.89\times 10^{56}\left(\frac{2.6\mu}{m}\right)^2\left(\frac{M_{halo}}{4M_{sun}}\right)^2\left(\frac{\ln 3.8}{\ln\frac{R_{out}}{R_{in}}}\right)^2 \frac{\left[1-\left(\frac{R_{in}}{R_{out}}\right)^3\right]}{0.98}[\text{cm}^{-3}] \end{aligned}\right\} \text{case I;} \tag{17}$$

$$\left.\begin{aligned} EM &= \frac{4}{5}\pi n_0^2 R_{in}^3 \left[1-\left(\frac{R_{in}}{R_{out}}\right)^5\right] \\ &= 7.86\times 10^{56}\left(\frac{2.6\mu}{m}\right)^2\left(\frac{M_{halo}}{4M_{sun}}\right)^2\left(\frac{0.74}{1-\frac{R_{in}}{R_{out}}}\right)^2 \frac{\left[1-\left(\frac{R_{in}}{R_{out}}\right)^5\right]}{0.99}[\text{cm}^{-3}] \end{aligned}\right\} \text{case II;} \tag{18}$$

$$\left.\begin{aligned} EM &= \frac{4}{5}\pi n_0^2 R_{in}^3 \left[1-\left(\frac{R_{in}}{R_{out}}\right)^5\right] \\ &= 1.82\times 10^{57}\left(\frac{2.6\mu}{m}\right)^2\left(\frac{M_{halo}}{4M_{sun}}\right)^2\left(\frac{0.47}{1-\frac{R_{in}}{R_{out}}}\right)^2 \frac{\left[1-\left(\frac{R_{in}}{R_{out}}\right)^5\right]}{0.96}[\text{cm}^{-3}] \end{aligned}\right\} \text{case III.} \tag{19}$$

Therefore we find the final expressions of the luminosity for Hβ

$$L(\text{H}\beta) = 1.43\times 10^{31}\left(\frac{T}{2.9\times 10^4}\right)^{-1.20}\left(\frac{2.6\mu}{m}\right)^2\left(\frac{M_{halo}}{4M_{sun}}\right)^2 \times$$

$$\left(\frac{\ln 3.8}{\ln\frac{R_{out}}{R_{in}}}\right)^2 \frac{\left[1-\left(\frac{R_{in}}{R_{out}}\right)^3\right]}{0.98}[\text{erg s}^{-1}], \quad \text{case I;} \tag{20}$$

$$L(\text{H}\beta) = 4.55\times 10^{31}\left(\frac{T}{2.3\times 10^4}\right)^{-0.833}\left(\frac{2.6\mu}{m}\right)^2\left(\frac{M_{halo}}{4M_{sun}}\right)^2$$

$$\left(\frac{0.74}{1-\frac{R_{in}}{R_{out}}}\right)^2 \frac{\left[1-\left(\frac{R_{in}}{R_{out}}\right)^5\right]}{0.99} \text{[erg s}^{-1}\text{]}, \quad \text{case II;} \tag{21}$$

$$L(H\beta) = 1.20\times 10^{32} \left(\frac{T}{2\times 10^4}\right)^{-0.833} \left(\frac{2.6\mu}{m}\right)^2 \left(\frac{M_{halo}}{4M_{sun}}\right)^2$$

$$\left(\frac{0.47}{1-\frac{R_{in}}{R_{out}}}\right)^2 \frac{\left[1-\left(\frac{R_{in}}{R_{out}}\right)^5\right]}{0.96} \text{[erg s}^{-1}\text{]}, \quad \text{case III;} \tag{22}$$

where we suppose the temperature to be in the neighborhood of $2.9\times 10^4$ K for case I, $2.3\times 10^4$ K for case II, and $2\times 10^4$ K for case III as computed below, and use the temperature power-law fit to $\frac{4\pi j_{H\beta}}{n_e n_p}$ given by Ferland (1980). This is approximate due to the assumption of Case B H I emission. We show below that the Lyman lines are not optically thick, and that continuum fluorescent excitation is important.

## 2.6. Scale radius

We can convert emission-line luminosities into surface brightness by dividing the luminosity by the area of emission on the sky. We assume that the lines form over a scale height determined by an effective radius, $R_{eff}$. The effective radius is defined as the position where half of the total line luminosity is formed. Emission line luminosities are determined by the emission measure, $n^2 V$ (AGN3), so the inner highest-density regions are most important. We obtain the effective or "half luminosity" radius from

$$\int_{R_{in}}^{R_{eff}} \frac{4\pi j}{n_e n_p} n(r)^2 \, dV = L/2 \text{[erg s}^{-1}\text{]}. \tag{23}$$

If we move $\frac{4\pi j_{H\beta}}{n_e n_p}$ out of the integral, equivalent to assuming that the temperature is constant, we find

$$R_{eff} = R_{in} \left( \dfrac{2}{1+\left(\dfrac{R_{in}}{R_{out}}\right)^3} \right)^{\frac{1}{3}}$$

$$= 6.26 \times 10^{18} \dfrac{\left( \dfrac{2}{1+\left(\dfrac{R_{in}}{R_{out}}\right)^3} \right)^{\frac{1}{3}}}{1.25} \text{[cm]}$$

$$= 1.25 R_{in}$$

$\left.\begin{array}{l}\end{array}\right\}$ case I and case II; (24)

$$R_{eff} = R_{in} \left( \dfrac{2}{1+\left(\dfrac{R_{in}}{R_{out}}\right)^5} \right)^{\frac{1}{5}}$$

$$= 5.67 \times 10^{18} \dfrac{\left( \dfrac{2}{1+\left(\dfrac{R_{in}}{R_{out}}\right)^5} \right)^{\frac{1}{5}}}{1.13} \text{[cm]}$$

$$= 1.15 R_{in}$$

$\left.\begin{array}{l}\end{array}\right\}$ case III. (25)

## 2.7. Average surface brightness in H I recombination lines

We convert the luminosities given above into surface brightness averaged over the full outer shell as it would be seen projected on the sky, in order to compare the results with observations. We obtain the surface brightness

$$S(H\beta) = \dfrac{1}{k^2} \dfrac{L}{4\pi^2 R_{eff}^2} \; [\text{erg s}^{-1} \text{ cm}^{-2} \text{ arcsec}^{-2}] \tag{26}$$

corresponding to

$$S(H\beta) = 2.76 \times 10^{-19} \left(\dfrac{T}{2.9 \times 10^4}\right)^{-1.20} \left(\dfrac{2.6\mu}{m}\right)^2 \left(\dfrac{M_{halo}}{4M_{sun}}\right)^2 \left(\dfrac{ln 3.8}{ln \dfrac{R_{out}}{R_{in}}}\right)^2 \times$$

$$S(H\beta) = \frac{\left[1-\left(\frac{R_{in}}{R_{out}}\right)^3\right]}{0.98} \frac{1.25}{\left[\frac{2}{1+\left(\frac{R_{in}}{R_{out}}\right)^3}\right]^{\frac{1}{3}}} \text{ [erg s}^{-1}\text{ cm}^{-2}\text{ arcsec}^{-2}\text{]}, \quad \text{case I;} \tag{27}$$

$$S(H\beta) = 9.42\times10^{-19}\left(\frac{T}{2.3\times10^4}\right)^{-0.833}\left(\frac{2.6\mu}{m}\right)^2\left(\frac{M_{halo}}{4M_{sun}}\right)^2\left(\frac{0.74}{1-\frac{R_{in}}{R_{out}}}\right)^2 \times$$

$$\frac{\left[1-\left(\frac{R_{in}}{R_{out}}\right)^5\right]}{0.99} \frac{1.15}{\left[\frac{2}{1+\left(\frac{R_{in}}{R_{out}}\right)^5}\right]^{\frac{1}{5}}} \text{ [erg s}^{-1}\text{ cm}^{-2}\text{ arcsec}^{-2}\text{]}, \quad \text{case II;} \tag{28}$$

and

$$S(H\beta) = 2.17\times10^{-18}\left(\frac{T}{2\times10^4}\right)^{-0.833}\left(\frac{2.6\mu}{m}\right)^2\left(\frac{M_{halo}}{4M_{sun}}\right)^2\left(\frac{0.47}{1-\frac{R_{in}}{R_{out}}}\right)^2 \times$$

$$\frac{\left[1-\left(\frac{R_{in}}{R_{out}}\right)^5\right]}{0.96} \frac{1.15}{\left[\frac{2}{1+\left(\frac{R_{in}}{R_{out}}\right)^5}\right]^{\frac{1}{5}}} \text{ [erg s}^{-1}\text{ cm}^{-2}\text{ arcsec}^{-2}\text{]}, \quad \text{case III;} \tag{29}$$

where $k = 206265$ converts luminosity into surface brightness. Table 5 in Sollerman et al. (2000) and Tziamtzis et al. (2009) also gave the surface brightness of the outer shell for different density laws.

The upper limit to the H$\beta$ surface brightness corresponds to an upper limit to the mass in the shell, for a given power-law index. The composition also affects $S(H\beta)$ because, for Crab abundances, the heavy elements contribute to the total mass. This means that the hydrogen density and $S(H\beta)$ are lower for the same mass but higher $Z$. The surface brightness is highest for solar abundances, where more of the $4M_{sun}$ is H so the density is higher. The coefficients in Equations (27), (28) and (29) were evaluated for abundances intermediate between solar and Crab. The maximum expansion velocity also affects the surface brightness because this sets the outer radius that appears in the equations. A shell with a larger expansion velocity is more spread out, has lower density, and lower $S(H\beta)$.

With these assumptions the physical conditions in the outer shell, the ionization and temperature, can be computed. Observations described below suggest that the upper limit to Hβ is about $S(Hβ) < 4\times10^{-18}$ erg cm$^{-2}$ s$^{-1}$ arcsec$^{-2}$. If we apply different values of $m$, indicating different abundances, into Equations (27), (28) and (29), we obtain that case I and case II have average surface brightness that are less than this observed limit for all abundances. Case III has a surface brightness that is under this observed limit for Crab abundances but above this observed limit for solar and ISM abundances. We consider all these models in the following to examine their predictions.

# 3. Model calculations

Here we will consider models with various compositions and power laws to compute the emitted spectrum. Case I and case II are more consistent with the existence of a large mass, $4M_{sun}$, and the limits to the surface brightness (Fesen, Shull & Hurford 1997; Tziamtzis et al. 2009). Since they have similar results for all kinds of calculations, we only give the full results for case I as examples. Sollerman et al. (2000) say that solar abundances might be most appropriate if the outer shell comes from the upper envelope of the star. We adopt this and further assume that grains have not formed in the fast wind. We present results for all the scenarios below, but will focus on this single model.

## 3.1. The emission-line spectrum

We use version 13 of the plasma simulation code Cloudy (Ferland et al. 2013) to predict the observed spectrum. We computed the luminosities of many emission lines and converted them to surface brightness by dividing the luminosities by the size derived above. We obtain different emission lines and surface brightness for the three different models. Figure 1 shows predicted spectra integrated over the full outer shell for case I with solar abundances. The upper panel shows the full range 0.1 to 100 microns. The lower panel shows the range 1 to 30 microns in greater detail. We focus on the UV – IR spectral region since this would be easiest to study with today's instrumentation.

Table 2 to Table 10 give the average surface brightnesses $S$ for IR, optical and UV emission lines, as defined by Equation (27), (28) and (29), for models with different abundances. These can be compared to the best observational upper limit achieved to date for the outer shell, $S < 1.2\times10^{-17}$ erg cm$^{-2}$ s$^{-1}$ arcsec$^{-2}$ using long-slit spectra to search for Hα (Fesen, Shull & Hurford 1997, but with their value corrected upwards by a factor of 3.4 to correct for the observed extinction ; Tziamtzis et al. 2009). All lines at or brighter than that limit are italicized in Table 2 to Table 10.

Hα is brighter than the observational limit in case III with solar or ISM

abundances, so at face value these models appear to be ruled out, at least for an outer shell containing the full amount of the missing mass. However, scattered light from the much brighter parts of the Crab is a major issue, as has been discussed by Tziamtzis et al. (2009). The Fesen, Shull & Hurford (1997) upper limit really corresponds to radii beyond about 0.3' from the bright edge of the main nebula ($R_{in}$), because their spectrum inside that radius is likely to be dominated by an unknown amount of scattered light. Figure 2 shows an example of the emissivity in three emission lines as a function of the depth into the shell from its inner edge at $R_{in}$. Figure 3 shows the results of integrating these emissivities along the line of sight through the outer shell to find the predicted surface brightness as a function of $R_{proj}/R_{in}$ for cases I, II and III. We used solar abundances and assumed a distance of 2 kpc and a spherical shell. Note that Figure 3 is shown with linear scales for both radius and surface brightness, to make the wide range in surface brightness more obvious, and that each panel has been separately scaled in surface brightness. Each panel also shows the Fesen, Shull & Hurford (1997) H$\alpha$ upper limit as a horizontal line beginning at a point 0.3' beyond $R_{in}$. The lack of an H$\alpha$ detection does nothing to rule out cases I or II, nor does it firmly rule out case III. Further ground-based observations might be able to push these optical-passband limits slightly fainter, but observations in H$\alpha$ or other lines that are also emitted by the main part of the nebula will require great attention to the scattered light issue.

What is needed are unique spectroscopic tracers in the form of high-ionization lines not emitted by the filaments or (hopefully) by the thin [O III]-emitting skin (Sankrit & Hester 1997) that surrounds the outer edge of the synchrotron bubble. Our models predict strong emission lines from high-ionization species of C, N, O, Ne, Si, Mg and Fe, principally in the UV and IR parts of the spectrum. A surface brightness limit similar to that for H$\alpha$ might be reachable in a few UV lines, notably C IV $\lambda\lambda$1548,1551, in about 3 hrs of on-target exposure (plus an equal sky exposure) using Hubble Space Telescope Advanced Camera for Surveys (ACS) imaging with very heavy on-detector binning. But the most promising lines are in the mid-IR, particularly [Ne VI] $\lambda$7.652µm which is also shown on Figure 3. Although these IR lines are somewhat fainter than the UV lines, they could be targeted with either SOFIA or (eventually) JWST mid-IR imagers and spectrographs. Archival Spitzer images and long-slit spectra also exist, and might be worth co-adding to search for these lines. Firm statements could be made about cases II or III if an IR measurement as deep as the H$\alpha$ limit could be obtained.

An alternative to searching areas off to the side of the main part of the Crab would be to obtain spectra averaging over a fairly large area at the center of the Crab where the projected expansion velocities are towards and away from us, and searching for these mid-IR lines with positive and negative velocity shifts corresponding to the shell structure. Lundqvist & Tziamtzis (2012) used this method in the optical passband to search for [O III] and Ca II lines. The [Ne VI] 7.652μm line falls within the spectral range covered by the Spitzer IRS, but Temim et al. (2012) do not report any strong feature at this wavelength in their IRS spectra of the Crab. The predicted spectral signature for such emission lines would be two broad peaks displaced symmetrically around the Crab's heliocentric systemic velocity of about 0 km s$^{-1}$ and separated by about 7500 km s$^{-1}$. We are in the process of carrying out the very careful reanalysis of the Spitzer spectra needed to search for faint features of this type. However, the low velocity resolution (4700 km s$^{-1}$) may prevent a clear distinction between any emission from an outer shell and emission from the ionized outer skin of the main part of the Crab (see Lundqvist & Tziamtzis 2012, their figure 9).

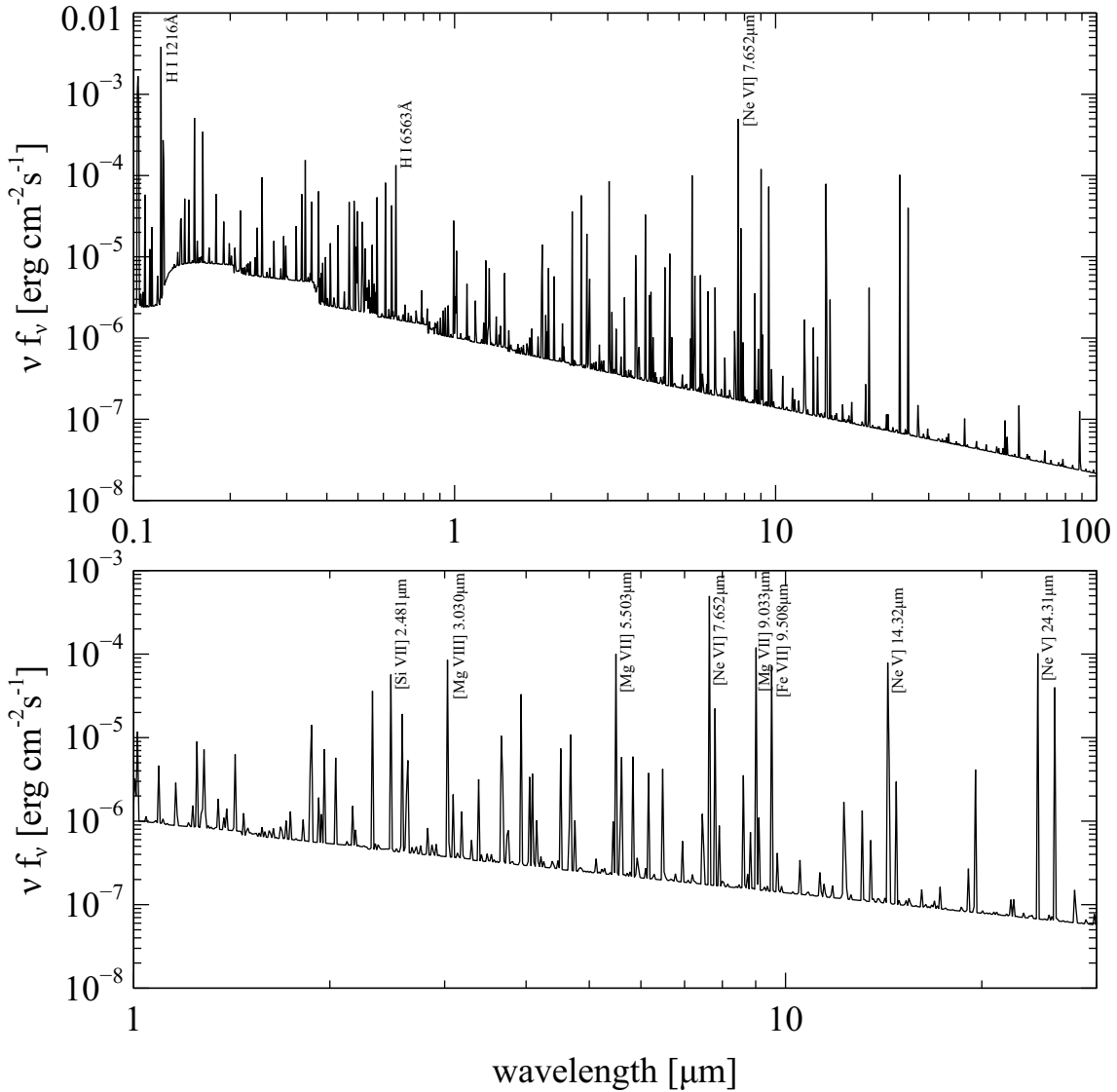

Figure 1. The upper panel shows emission lines from Crab outer shell between the wavelength of 0.1μm and 100 μm for case I with solar abundances. H I 1216Å, H I 6563Å and [Ne VI] 7.652 μm are the strongest lines in UV, optical and IR bands respectively. The lower panel shows the emission lines for the same model in the range between 1 μm and 30 μm and all the lines that are brighter than H$\beta$ line are marked on the figure.

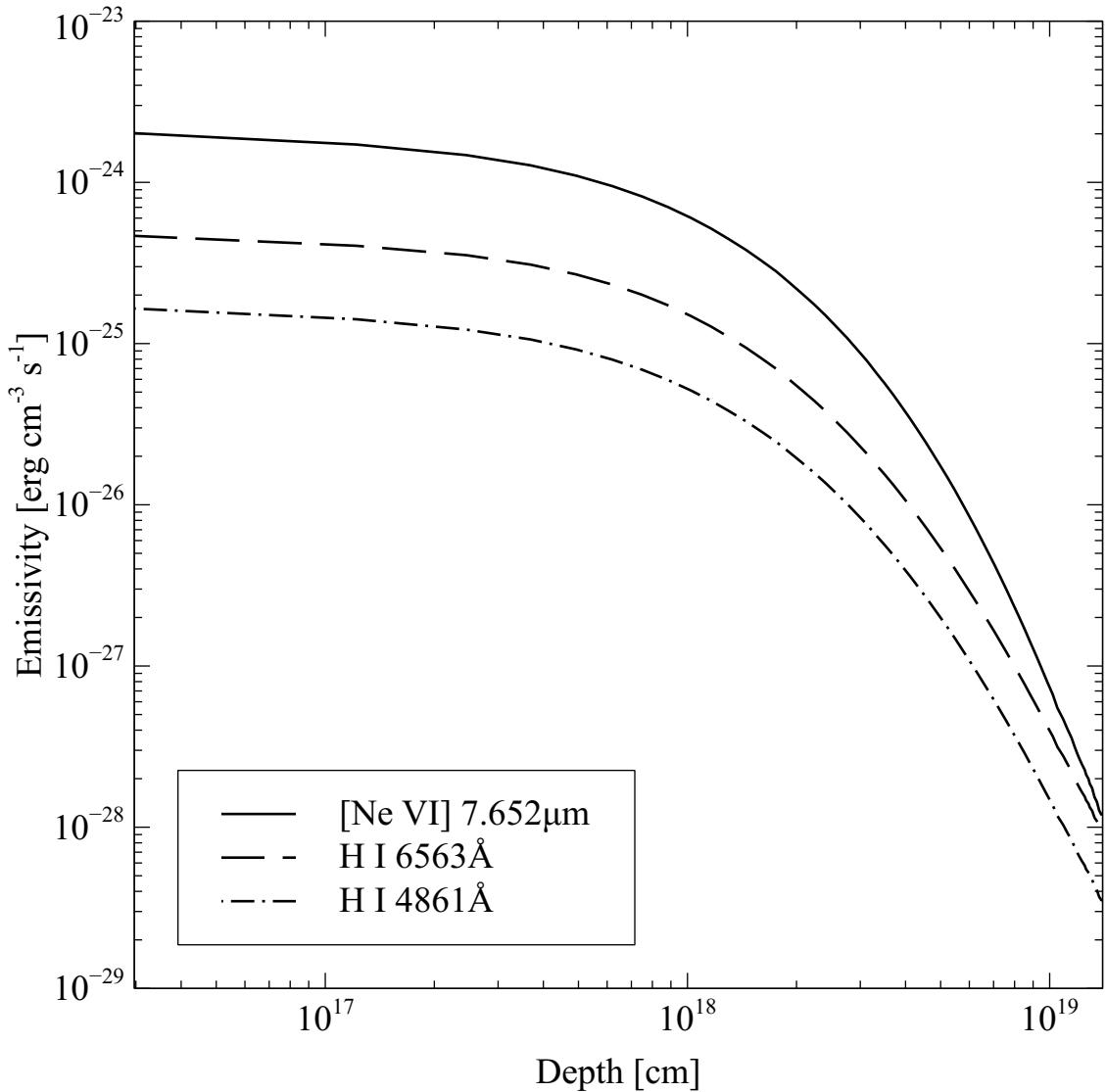

Figure 2. Emissivity as a function of depth of lines [Ne VI] 7.652$\mu$m, H I 6563Å and H I 4861Å, for case I. For a distance of 2kpc, $2\times10^{18}$ cm corresponds to 1.1'.

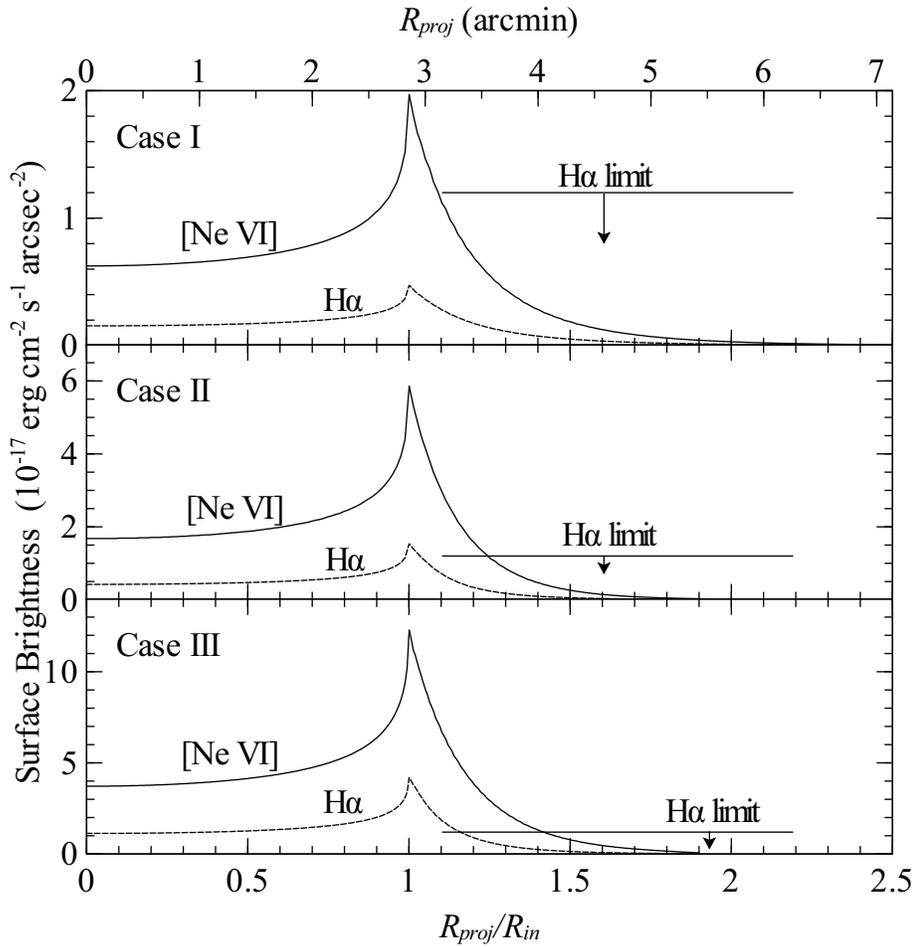

Figure 3. Predicted surface brightness of the [Ne VI] λ7.652μm and Hα emission lines, as a function of $R_{proj}$, the radial distance from the center of expansion as seen projected on the sky. These are computed for Cases I, II and III with solar abundances and assuming a distance of 2 kpc and a spherical outer shell of inner radius $R_{in} = 5 \times 10^{18}$ cm. The surface brightness for $R_{proj} < R_{in}$ includes both the front and rear sides of the outer shell. The horizontal bar in each panel shows the Fesen, Shull & Hurford (1997) Hα upper limit discussed in the text, starting at a point 0.3' beyond $R_{in}$ and extending to the end of their slit.

Cloudy predicts the intensity of H I lines including line optical depths effect, collisional excitation and de-excitation, and continuum fluorescent excitation. The predicted H I intensities can be compared with Case B (pure recombination in which Lyman lines are optically thick) and Case A (Lyman lines are optically thin and there is no continuum fluorescent excitation).

Table 11 compares H I luminosities for the solar abundance, case I Crab shell. It gives the computed luminosities with all processes included, along with the luminosities obtained from the computed density and temperature and assuming Case A and Case B emission (Storey & Hummer 1995). The predicted lines are about 10%~140% brighter than Case B, an indication that continuum fluorescent excitation is important. The Lyman lines in the outer shell are not optically thick so continuum pumping is important, causing them to be brighter than would be found with pure recombination. The optical depth in Ly$\beta$, for instance, is about 1, so neither Case A nor Case B formally apply. The predicted deviations are not large and Case B is, as is often the case, a fair approximation to the actual emission.

### 3.2. Gas Temperature

Figure 4 shows the gas kinetic temperature across the outer shell. It increases as the depth increases for all three models. This is because the Crab radiation field, which powers the outer shell, decreases at $r^{-2}$, because of the inverse square law. The gas density falls off faster, as $r^{-3}$ or $r^{-4}$. As a result the ionization parameter, the ratio of photon to hydrogen densities (AGN3), increases as *r* increases. Higher ionization parameter gas tends to be hotter.

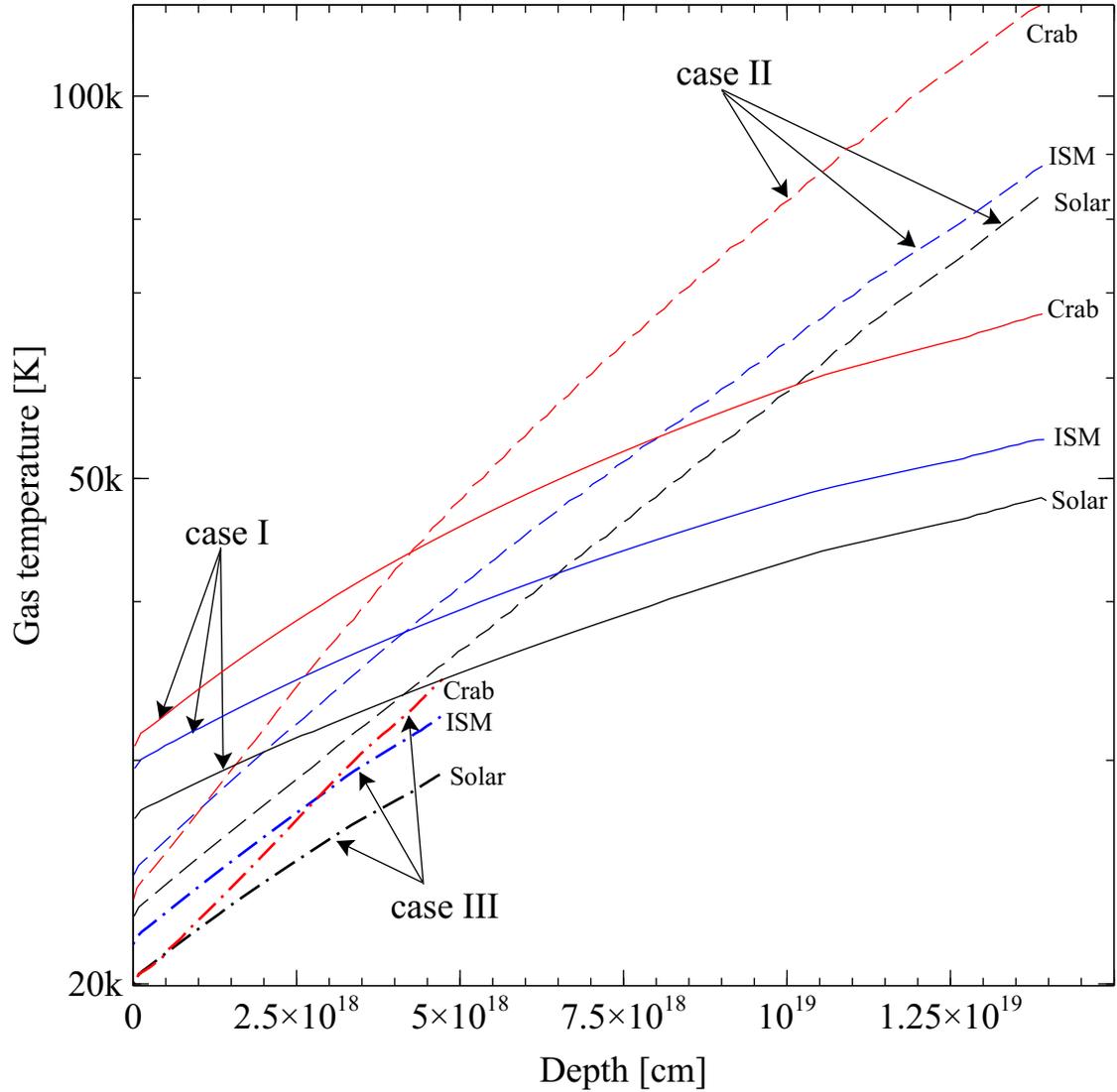

Figure 4. Gas temperature across the Crab outer shell for all the models. The depth is the distance between the illuminated face of the outer shell and a point within the outer shell.

### 3.3. The absorption line spectrum

We compute optical depths for different assumptions about the expansion velocities. Table 12 to Table 14 give the optical depths for models with Crab abundances, solar abundances and ISM abundances respectively. We continue to focus on the models with solar abundances. From Table 12 to Table 14 we see that the optical depths for C IV 1549 doublet are not much greater than 1. The lines mainly form over a small radius due to the density decline, so the wind acceleration should not be large over the line forming region.

We make two assumptions to estimate the optical depth. First we assume a static shell. The lines are only thermally broadened. This would apply if there is no acceleration across the layer where the lines form. In this case there is sufficient opacity to produce the observed lines. In particular, the C IV 1549 doublet has an optical depth of 2.68, consistent with the Sollerman et al. (2000) tentative detection. We note that the optical depth of the O VI 1034 doublet is much larger than 1, which indicates strong absorption at that wavelength.

If the lines have a significant component of turbulence or if the expansion velocity changes across the line-forming region then the lines will be spread over a wider velocity range. Here the lines are optically thin.

Table 15 gives the computed line optical depths for both static and dynamic cases. We find the optical depth to be very small if we add a turbulence with velocity $v$ =1680 km s$^{-1}$ as the expansion velocity of the inner radius of the outer shell. The O VI 1034 doublet becomes optically thin as well.

The truth will lie between these two limiting assumptions. We will consider dynamic models, in which the velocity is determined self consistently, in future papers.

### 3.4. Is steady state appropriate?

#### Recombination time scale

The recombination time scale is defined as (AGN3)

$$t_{rec} = \frac{1}{n_e \alpha_B(T_e)} \quad (30)$$

where $n_e$ is the electron density and $\alpha_B(T_e)$ is the Case B recombination coefficient at temperature $T_e$. The gas in the outer shell is photoionized by light from the visible Crab. Since Cloudy supposes that the gas atomic processes that are responsible for thermal and ionization equilibrium have reached steady state, we need to compare the age of the Crab with the recombination time to see if this is valid. We compute the recombination timescale for Ne$^{+6}$ → Ne$^{+5}$ for all three cases with solar abundances. We focus on this ion since it produces the strongest IR line. Since the temperature increases very slowly but the electron density decreases very quickly, we assume different radii have roughly the same recombination coefficient and evaluate it from the Badnell (2006), Badnell et al. (2003) and Badnell web site (http://amdpp.phys.strath.ac.uk/tamoc/DR/). We find the recombination time are about 100 years, 20 years and 10 years in the inner edge of the shell for the case I, case II and case III respectively (Figure 5). All of these are much shorter than the age of the visible Crab, suggesting that the outer shell has reached photoionization equilibrium.

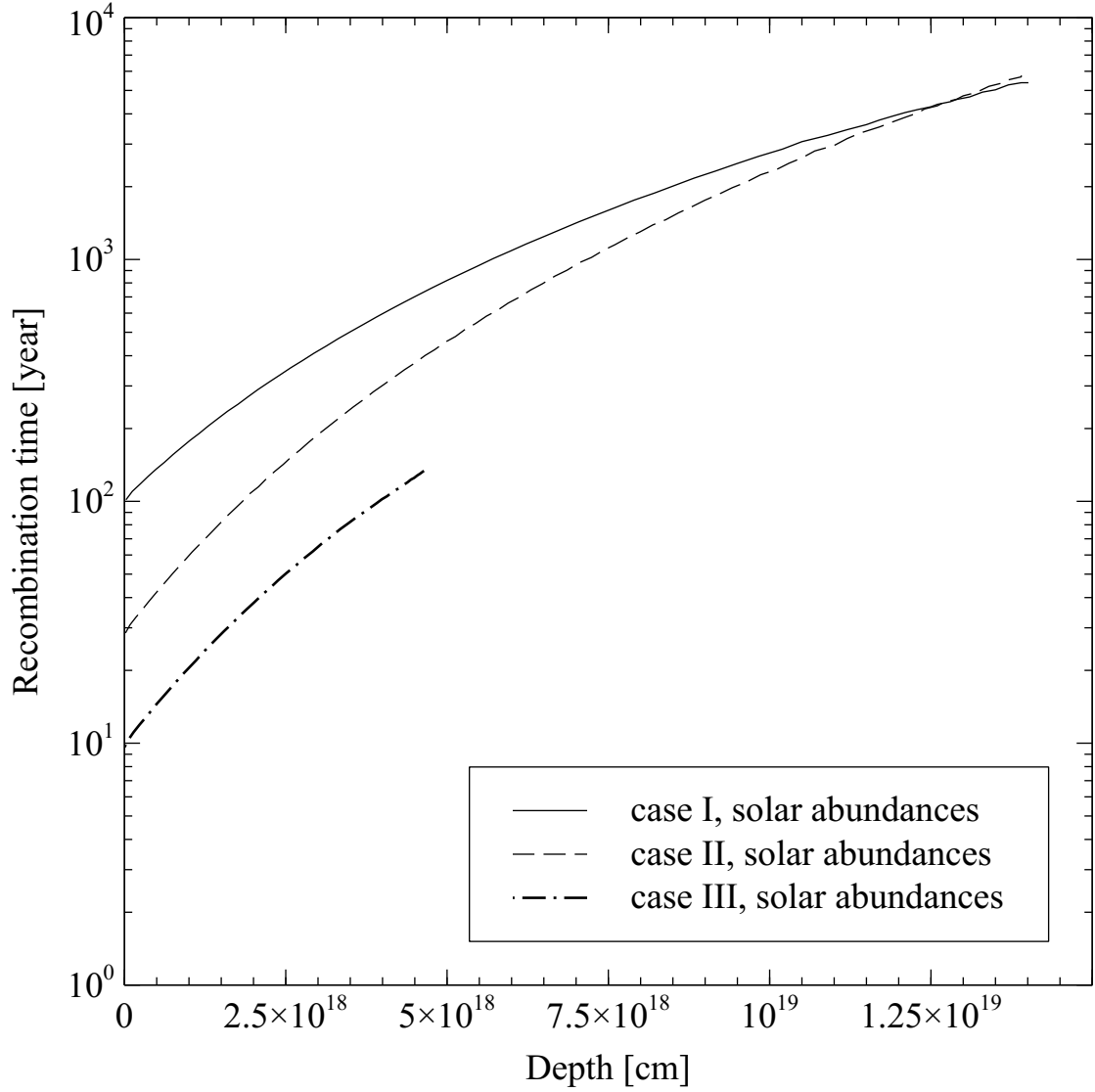

Figure 5. Recombination time scales for producing Ne$^{+5}$ as a function of depth in the Crab outer shell for all three cases with solar abundances.

### Thermal timescale

We calculate both the thermal energy [erg cm$^{-3}$] and the cooling rate [erg cm$^{-3}$s$^{-1}$] as a function of the radius for all three cases with solar abundances. From the ratio we can find the cooling time. We also calculate the emission measure for different radii or different zones. The differential emission measure for each depth is then

$$dEM = 4\pi r^2 n(r)^2 dr. \tag{31}$$

This gives an indication of which portions of the shell contribute most to the observed emission.

Figure 6 shows the cooling times and the differential emission measure across the Crab outer shell for all three cases with solar abundances. We find the cooling time for all cases to be much longer than the age of the visible Crab. Even for the inner edge, which produces much of the emission measure, the cooling times are still about 20, 10 and 6 times of the age of the visible Crab for case I, case II and case III respectively. This indicates that the outer shell has not had time to reach thermal equilibrium, so retains a memory of its temperature in the past.

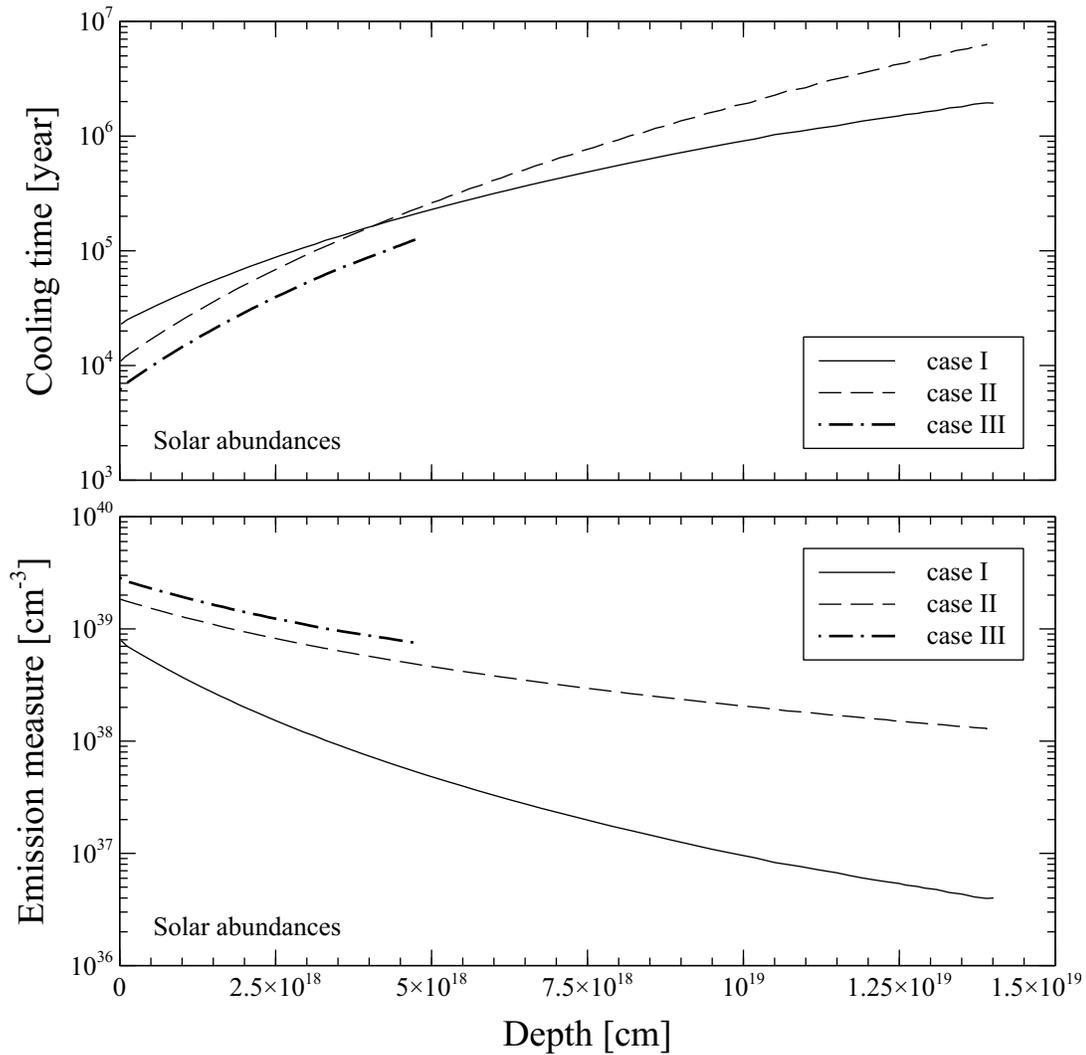

Figure 6. Cooling times and differential emission measures for all three cases with solar abundances. The innermost regions have the greatest emission measure and so would contribute the most to the observed spectrum.

### Effects on predicted spectrum

The shell is in photoionization, but is not in thermal, equilibrium. This means that atomic processes which set the ionization of the gas have reached steady state, and that the predicted ionization should be accurate. The fact that the gas is not in thermal equilibrium means that we don't really know its temperature, only that it is young enough to "remember" its temperature long ago. In other words, the current temperature is partially determined by its temperature in the past. We don't know whether the outer shell was initially hot or cold.

All of this is important because we predict that high ionization IR lines should be among the strongest lines in the optical-IR spectrum. Are these predictions approximately valid?

The uncertain temperature should not greatly affect the high ionization lines in the IR. The emissivities of an IR collisionally excited line do not have a strong temperature dependence. The lines have low excitation potentials, their Boltzmann factors should be close to unity, so their emissivity is proportional to $T^{-\frac{1}{2}}$ (AGN3). The optical recombination lines have an emissivity that is a faster power law, typically $T^{-0.8}$. Factors of two uncertainties in the temperature carry over to uncertainties in the line's surface brightness by well less than a factor of two.

Similarly, the uncertain temperature should not greatly affect the predicted ionization of the gas. The ionization is set by the photoionization and recombination rates. The photoionization rate has no temperature dependence, while recombination coefficients have power-law temperature dependencies, roughly $T^{-0.8}$. Factors of two uncertainties in the temperature will change the ionization by less than this.

Lundqvist, Fransson, & Chevalier (1986) gave time-dependent numerical simulations. We do have the ability to do time dependent, fully advective, photoionization flows (Henney et al. 2005; 2007). However these calculations would have to be guided by observations that do not now exist. Is the shell cooling down from a hotter phase, warming up from a colder phase, or is it now in approximate thermal equilibrium?

## 4. Discussion and conclusions

We have presented a series of photoionization equilibrium calculations of the properties of the outer shell in the Crab Nebula. We reach the following conclusions.

- The gas cooling time is far longer than age of visible Crab, so the outer shell is not in thermal equilibrium. As a result we don't really know its temperature since it will carry a memory of its original value.
- The recombination time is much shorter than the age of the Crab, so the outer shell is in ionization equilibrium.
- Together these mean that the outer shell will be highly ionized but we are not certain of its temperature. We find that the IR coronal lines are very strong, stronger than most optical lines used in previous searches. Luckily, these lines are not sensitive to the gas temperature so this is a robust prediction.
- The outer shell can produce the observed C IV absorption if the line broadening across the line-forming region is not large. Full dynamical solutions would be needed to make robust predictions of this line optical depth.
- The existing observational limit on H$\alpha$ does not place useful constraints on most of our models, but is on the verge of ruling out models with solar and ISM abundances and $\alpha < -4$, $v \propto r^2$ and containing the full amount of the missing mass.
- The IR coronal lines are our best hope for avoiding confusion with scattered light from the inner parts of the Crab. The species producing them are too highly ionized to be produced by the photoionized gas in the filaments, and are higher ionization than the shocked gas that directly produces the [O III] emission skin at the outer edge of the synchrotron bubble (although higher velocity shocks could co-exist in this latter region and produce such lines).
- We recommend imaging (or spectroscopy) on the sky just outside the main part of the Crab to search for one of these IR lines.
- An alternative approach would be to search for these lines in spectra of the center of the Crab where the projected expansion velocities are towards and away from us.

We thank Peter Lundqvist for his careful review of our manuscript. GJF acknowledges support by NSF (1108928; and 1109061), NASA (10-ATP10-0053, 10-ADAP10-0073, and NNX12AH73G), and STScI (HST-AR-12125.01, GO-12560, and HST-GO-12309). JAB and CTR acknowledge support by NSF (1006593). CTR, JAB, and EDL are grateful to NASA for support through ADAP grant NNX10AC93G.

**Table 1. Basic parameters of the outer shell for three difference cases**

| Case | $\gamma$ | $\alpha$ | $\beta$ | $n_0$(cm$^{-3}$) | $R_{out}$(cm) |
|------|----------|----------|---------|------------------|---------------|
| I    | 1        | -3       | 0       | 0.87             | 1.90E+19      |
| II   | 1        | -4       | 0       | 1.58             | 1.90E+19      |
| III  | 2        | -4       | -1      | 2.46             | 9.50E+18      |

**Table 2. Predicted IR emission line average surface brightness, sorted by surface brightness for each model, for all lines brighter than H$\beta$, for case I.**

| Crab Abund. | | Solar Abund. | | ISM Abund. | |
|---|---|---|---|---|---|
| Line[1] | Surf. Br.[2] | Line | Surf. Br. | Line | Surf. Br. |
| *[Ne VI] 7.652m* | *1E-17* | *[Ne VI] 7.652m* | *1E-17* | *[Ne VI] 7.652m* | *1E-17* |
| [Ne V] 24.31m | 2E-18 | [Mg VII] 9.033m | 3E-18 | [Ne V] 24.31m | 3E-18 |
| [Mg VIII] 3.030m | 2E-18 | [Ne V] 24.31m | 3E-18 | [Ne V] 14.32m | 2E-18 |
| [Mg VII] 9.033m | 2E-18 | [Mg VII] 5.503m | 3E-18 | [S VIII] 9914 | 1E-18 |
| [Mg VII] 5.503m | 2E-18 | [Mg VIII] 3.030m | 2E-18 | [Mg VII] 9.033m | 1E-18 |
| [Ne V] 14.32m | 2E-18 | [Ne V] 14.32m | 2E-18 | | |
| He II 1.012m | 5E-19 | [Fe VII] 9.508m | 2E-18 | | |
| [O IV] 25.88m | 5E-19 | [Si VII] 2.481m | 1E-18 | | |
| [Fe VII] 9.508m | 4E-19 | | | | |
| [Si IX] 3.929m | 4E-19 | | | | |
| [S VIII] 9914 | 3E-19 | | | | |
| [Si VII] 2.481m | 3E-19 | | | | |

[1] Wavelengths are given in Å unless noted with m = microns.

[2] Surface brightness, erg cm$^{-2}$ s$^{-1}$ arcsec$^{-2}$.

**Table 3. Predicted optical emission line average surface brightness, sorted by surface brightness for each model, for all lines brighter than H$\beta$, for case I.**

| Crab Abund. | | Solar Abund. | | ISM Abund. | |
|---|---|---|---|---|---|
| Line[1] | Surf. Br.[2] | Line | Surf. Br. | Line | Surf. Br. |
| He II 4686 | 2E-18 | H I 6563 | 3E-18 | H I 6563 | 3E-18 |
| H I 6563 | 8E-19 | Fe VII 6087 | 2E-18 | H I 4861 | 1E-18 |
| Fe X 6375 | 7E-19 | Fe VII 5721 | 1E-18 | | |
| Fe VII 6087 | 5E-19 | H I 4861 | 1E-18 | | |
| Fe VII 5721 | 3E-19 | | | | |
| H I 4861 | 3E-19 | | | | |

[1] Wavelengths are given in Å unless noted with m = microns.

[2] Surface brightness, erg cm$^{-2}$ s$^{-1}$ arcsec$^{-2}$.

Italicized entries have predicted surface brightness at or above the current optical-passband detection limit.

**Table 4. Predicted UV emission line average surface brightness, sorted by surface brightness for each model, for all lines brighter than H$\beta$, for case I.**

| Crab Abund. | | solar Abund. | | ISM Abund. | |
|---|---|---|---|---|---|
| Line[1] | Surf. Br.[2] | Line | Surf. Br. | Line | Surf. Br. |
| *O VI 1032+1038* | *1E-16* | *O VI 1032+1038* | *2E-16* | *H I 1216* | *9E-17* |
| *C IV 1548+1551* | *1E-16* | *H I 1216* | *9E-17* | *O VI 1032+1038* | *9E-17* |
| *H I 1216* | *4E-17* | *H I 1026* | *3E-17* | *C IV 1548+1551* | *3E-17* |
| *He II 1640* | *1E-17* | *C IV 1548+1551* | *2E-17* | *H I 1026* | *3E-17* |
| *O V 1211+1218* | *1E-17* | *N V 1239+1243* | *2E-17* | *N V 1239+1243* | *2E-17* |
| *N V 1239+1243* | *1E-17* | *O V 1211+1218* | *1E-17* | *O V 1211+1218* | *1E-17* |
| H I 1026 | 9E-18 | He II 1640 | 9E-18 | He II 1640 | 8E-18 |
| He II 1215 | 5E-18 | [Ne V] 3426 | 3E-18 | [Ne V] 3426 | 4E-18 |
| [Ne V] 3426 | 3E-18 | He II 1215 | 3E-18 | He II 1215 | 3E-18 |
| He II 1085 | 2E-18 | Mg VII 2569 | 2E-18 | Ne V 3346 | 2E-18 |
| C III] 1907+1910 | 2E-18 | Fe VII 3759 | 2E-18 | He II 1085 | 1E-18 |
| Mg VII 2569 | 2E-18 | He II 1085 | 1E-18 | | |
| He II 1025 | 1E-18 | Ne V 3346 | 1E-18 | | |
| Ne V 3346 | 1E-18 | [Mg VI] 1806 | 1E-18 | | |
| C V 1312 | 1E-18 | | | | |
| He II 2050 | 1E-18 | | | | |
| [C V] 2271+2275 | 8E-19 | | | | |
| He II 3203 | 8E-19 | | | | |
| Ne V 1141 | 7E-19 | | | | |
| [Mg VI] 1806 | 7E-19 | | | | |
| Si VIII 1446 | 5E-19 | | | | |
| He II 2733 | 4E-19 | | | | |
| C VI 1240 | 4E-19 | | | | |
| Fe VII 3759 | 4E-19 | | | | |
| He II 3645 | 3E-19 | | | | |
| O IV 1405 | 3E-19 | | | | |
| [Fe VII] 3586 | 3E-19 | | | | |

[1] Wavelengths are given in Å unless noted with m = microns.

[2] Surface brightness, erg cm$^{-2}$ s$^{-1}$ arcsec$^{-2}$.

Italicized entries have predicted surface brightness at or above the current optical-passband detection limit.

**Table 5. Predicted IR emission line average surface brightness, sorted by surface brightness for each model, for all lines brighter than H$\beta$, for case II.**

| Crab Abund. | | Solar Abund. | | ISM Abund. | |
|---|---|---|---|---|---|
| Line[1] | Surf. Br.[2] | Line | Surf. Br. | Line | Surf. Br. |
| *[Ne VI] 7.652m* | *5E-17* | *[Ne VI] 7.652m* | *4E-17* | *[Ne VI] 7.652m* | *4E-17* |
| *[Ne V] 24.31m* | *1E-17* | *[Ne V] 24.31m* | *1E-17* | *[Ne V] 24.31m* | *2E-17* |
| *[Ne V] 14.32m* | *1E-17* | *[Ne V] 14.32m* | *1E-17* | *[Ne V] 14.32m* | *1E-17* |
| [O IV] 25.88m | 5E-18 | [O IV] 25.88m | 9E-18 | [O IV] 25.88m | 5E-18 |
| [Mg VII] 9.033m | 4E-18 | [Fe VII] 9.508m | 7E-18 | | |
| [Mg VII] 5.503m | 4E-18 | [Mg VII] 9.033m | 6E-18 | | |
| [Mg VIII] 3.030m | 3E-18 | [Mg VII] 5.503m | 5E-18 | | |
| [Fe VII] 9.508m | 2E-18 | [Si VII] 2.481m | 4E-18 | | |
| He II 1.012m | 1E-18 | | | | |
| [Si VII] 2.481m | 1E-18 | | | | |
| [S VIII] 9914 | 8E-19 | | | | |

[1] Wavelengths are given in Å unless noted with m = microns.

[2] Surface brightness, erg cm$^{-2}$ s$^{-1}$ arcsec$^{-2}$.

Italicized entries have predicted surface brightness at or above the current optical-passband detection limit.

**Table 6. Predicted optical emission line average surface brightness, sorted by surface brightness for each model, for all lines brighter than H$\beta$, for case II.**

| Crab Abund. | | solar Abund. | | ISM Abund. | |
|---|---|---|---|---|---|
| Line[1] | Surf. Br.[2] | Line | Surf. Br. | Line | Surf. Br. |
| He II 4686 | 6E-18 | H I 6563 | 9E-18 | H I 6563 | 9E-18 |
| H I 6563 | 2E-18 | Fe VII 6087 | 7E-18 | H I 4861 | 3E-18 |
| Fe VII 6087 | 2E-18 | Fe VII 5721 | 4E-18 | | |
| Fe VII 5721 | 1E-18 | H I 4861 | 3E-18 | | |
| H I 4861 | 8E-19 | | | | |

[1] Wavelengths are given in Å unless noted with m = microns.

[2] Surface brightness, erg cm$^{-2}$ s$^{-1}$ arcsec$^{-2}$.

Italicized entries have predicted surface brightness at or above the current optical-passband detection limit.

**Table 7. Predicted UV emission line average surface brightness, sorted by surface brightness for each model, for all lines brighter than H$\beta$, for case II.**

| Crab Abund. | | Solar Abund. | | ISM Abund. | |
| --- | --- | --- | --- | --- | --- |
| Line[1] | Surf. Br.[2] | Line | Surf. Br. | Line | Surf. Br. |
| C IV 1548+1551 | 3E-16 | H I 1216 | 2E-16 | H I 1216 | 2E-16 |
| O VI 1032+1038 | 2E-16 | O VI 1032+1038 | 1E-16 | O VI 1032+1038 | 1E-16 |
| H I 1216 | 8E-17 | C IV 1548+1551 | 7E-17 | C IV 1548+1551 | 8E-17 |
| He II 1640 | 4E-17 | N V 1239+1243 | 5E-17 | N V 1239+1243 | 5E-17 |
| O V 1211+1218 | 3E-17 | H I 1026 | 5E-17 | H I 1026 | 5E-17 |
| N V 1239+1243 | 2E-17 | O V 1211+1218 | 3E-17 | O V 1211+1218 | 3E-17 |
| H I 1026 | 2E-17 | He II 1640 | 2E-17 | He II 1640 | 2E-17 |
| Ne V 3426 | 2E-17 | Ne V 3426 | 1E-17 | Ne V 3426 | 2E-17 |
| He II 1215 | 1E-17 | He II 1215 | 8E-18 | He II 1215 | 8E-18 |
| C III 1907+1910 | 1E-17 | Ne V 3346 | 5E-18 | Ne V 3346 | 6E-18 |
| He II 1085 | 7E-18 | Fe VII 3759 | 4E-18 | He II 1085 | 4E-18 |
| Ne V 3346 | 6E-18 | He II 1085 | 4E-18 | | |
| HE 2 1025 | 4E-18 | | | | |
| Mg VII 2569 | 3E-18 | | | | |
| He II 3203 | 3E-18 | | | | |
| C V 1312 | 2E-18 | | | | |
| O IV 1401+1405 | 2E-18 | | | | |
| Ne IV 2424 | 2E-18 | | | | |
| Mg VI 1806 | 2E-18 | | | | |
| Ne V 1141 | 2E-18 | | | | |
| He II 2733 | 1E-18 | | | | |
| Fe VII 3759 | 1E-18 | | | | |
| C V 2275 | 1E-18 | | | | |
| Fe VII 3586 | 9E-19 | | | | |
| He II 2511 | 8E-19 | | | | |

[1] Wavelengths are given in Å unless noted with m = microns.

[2]Surface brightness, erg cm$^{-2}$ s$^{-1}$ arcsec$^{-2}$.

Italicized entries have predicted surface brightness at or above the current optical-passband detection limit.

**Table 8. Predicted IR emission line average surface brightness, sorted by surface brightness for each model, for all lines brighter than H$\beta$, for case III.**

| Crab Abund. | | Solar Abund. | | ISM Abund. | |
|---|---|---|---|---|---|
| Line[1] | Surf. Br.[2] | Line | Surf. Br. | Line | Surf. Br. |
| *[Ne VI] 7.652m* | *1E-16* | *[Ne VI] 7.652m* | *8E-17* | *[Ne VI] 7.652m* | *1E-16* |
| *[Ne V] 24.31m* | *7E-17* | *[Ne V] 24.31m* | *5E-17* | *[Ne V] 24.31m* | *6E-17* |
| *[Ne V] 14.32m* | *5E-17* | *[O IV] 25.88m* | *5E-17* | *[Ne V] 14.32m* | *5E-17* |
| *[O IV] 25.88m* | *3E-17* | *[Ne V] 14.32m* | *4E-17* | *[O IV] 25.88m* | *3E-17* |
| [Mg VII] 9.033m | 8E-18 | [Fe VII] 9.508m | 2E-17 | | |
| [Mg VII] 5.503m | 7E-18 | [Mg VII] 9.033m | 1E-17 | | |
| [Fe VII] 9.508m | 6E-18 | [Si VII] 2.481m | 1E-17 | | |
| [He II] 1.012m | 4E-18 | | | | |
| [Mg VIII] 3.030m | 3E-18 | | | | |
| [Si VII] 2.481m | 3E-18 | | | | |

[1] Wavelengths are given in Å unless noted with m = microns.

[2] Surface brightness, erg cm$^{-2}$ s$^{-1}$ arcsec$^{-2}$.

Italicized entries have predicted surface brightness at or above the current optical-passband detection limit.

**Table 9. Predicted optical emission line average surface brightness, sorted by surface brightness for each model, for all lines brighter than H$\beta$, for case III.**

| Crab Abund. | | Solar Abund. | | ISM Abund. | |
|---|---|---|---|---|---|
| Line[1] | Surf. Br.[2] | Line | Surf. Br. | Line | Surf. Br. |
| *He II 4686* | *2E-17* | *H I 6563* | *2E-17* | *H I 6563* | *2E-17* |
| H I 6563 | 6E-18 | *Fe VII 6087* | *2E-17* | H I 4861 | 8E-18 |
| Fe VII 6087 | 5E-18 | *Fe VII 5721* | *1E-17* | | |
| Fe VII 5721 | 3E-18 | H I 4861 | 9E-18 | | |
| H I 4861 | 2E-18 | | | | |

[1] Wavelengths are given in Å unless noted with m = microns.

[2]Surface brightness, erg cm$^{-2}$ s$^{-1}$ arcsec$^{-2}$.

Italicized entries have predicted surface brightness at or above the current optical-passband detection limit.

**Table 10. Predicted UV emission line average surface brightness, sorted by surface brightness for each model, for all lines brighter than H$\beta$, for case III.**

| Crab Abund. | | Solar Abund. | | ISM Abund. | |
|---|---|---|---|---|---|
| Line[1] | Surf. Br.[2] | Line | Surf. Br. | Line | Surf. Br. |
| *C IV 1548+1551* | *9E-16* | *H I 1216* | *3E-16* | *H I 1216* | *3E-16* |
| *O VI 1032+1038* | *2E-16* | *C IV 1548+1551* | *2E-16* | *C IV 1548+1551* | *2E-16* |
| *H I 1216* | *1E-16* | *O VI 1032+1038* | *2E-16* | *O VI 1032+1038* | *1E-16* |
| *He II 1640* | *1E-16* | *N V 1239+1243* | *8E-17* | *N V 1239+1243* | *9E-17* |
| *C III] 1907+1910* | *5E-17* | *He II 1640* | *6E-17* | *H I 1026* | *6E-17* |
| *[Ne V] 3426* | *5E-17* | *H I 1026* | *6E-17* | *He II 1640* | *6E-17* |
| *O V 1211+1218* | *5E-17* | *O V 1211+1218* | *6E-17* | *[Ne V] 3426* | *5E-17* |
| *He II 1215* | *4E-17* | *[Ne V] 3426* | *4E-17* | *O V 1211+1218* | *5E-17* |
| *N V 1239+1243* | *4E-17* | *He II 1215* | *2E-17* | *He II 1215* | *2E-17* |
| *H I 1026* | *4E-17* | *Ne V 3346* | *1E-17* | *Ne V 3346* | *2E-17* |
| *He II 1085* | *2E-17* | *Fe VII 3759* | *1E-17* | *Ne IV 2424* | *1E-17* |
| *Ne V 3346* | *2E-17* | *He II 1085* | *1E-17* | *N IV 1485* | *1E-17* |
| *He II 1025* | *1E-17* | *Ne IV 2424* | *1E-17* | He II 1085 | 9E-18 |
| Ne IV 2424 | 9E-18 | | | C III] 1907 | 9E-18 |
| He II 3203 | 7E-18 | | | | |
| He II 2050 | 7E-18 | | | | |
| C V 1312 | 4E-18 | | | | |
| He II 2733 | 4E-18 | | | | |
| O IV 1405 | 4E-18 | | | | |
| Mg VII 2569 | 3E-18 | | | | |
| [Mg VI] 1806 | 3E-18 | | | | |
| Fe VII 3759 | 3E-18 | | | | |

| Line | Surface Brightness |
|---|---|
| Ne V 1141 | 3E-18 |
| O IV 1401 | 3E-18 |
| Mg V 2855 | 3E-18 |
| He II 2511 | 2E-18 |

[1] Wavelengths are given in Å unless noted with m = microns.

[2] Surface brightness, erg cm$^{-2}$ s$^{-1}$ arcsec$^{-2}$.

Italicized entries have predicted surface brightness at or above the current optical-passband detection limit.

**Table 11. H I luminosities for different cases (case I, solar) [erg s$^{-1}$]**

| Line | Total | Case B | Case A |
|---|---|---|---|
| H I 6563Å | 2.09E+32 | 1.75E+32 | 1.04E+32 |
| H I 4861Å | 7.31E+31 | 6.47E+31 | 4.02E+31 |
| H I 1216Å | 5.86E+33 | 2.37E+33 | 1.40E+33 |
| H I 1.875$\mu$m | 2.05E+31 | 1.68E+31 | 1.43E+31 |

**Table 12. Predicted optical depth, sorted by wavelength for thermal-broadened model, for case I.**

| Crab Abund. | | Solar Abund. | | ISM Abund. | |
|---|---|---|---|---|---|
| Line[1] | Opt. Dpt.[2] | Line | Opt. Dpt. | Line | Opt. Dpt. |
| O I 1025 | 3.56E-01 | O I 1025 | 1.62E+00 | O I 1025 | 1.42E+00 |
| H I 1025 | 4.05E-01 | H I 1025 | 1.88E+00 | H I 1025 | 1.63E+00 |
| O VI 1031+1037 | 3.21E+01 | O VI 1031+1037 | 3.81E+01 | O VI 1031+1037 | 2.19E+01 |
| H I 1215 | 2.53E+00 | H I 1215 | 1.17E+01 | H I 1215 | 1.02E+01 |

| Line | Opt. Dpt. | Line | Opt. Dpt. | Line | Opt. Dpt. |
|---|---|---|---|---|---|
| N V 1239+1243 | 1.03E+00 | N V 1239+1243 | 3.54E+00 | N V 1239+1243 | 2.88E+00 |
| C IV 1548+1551 | 1.32E+01 | C IV 1548+1551 | 2.68E+00 | C IV 1548+1551 | 2.33E+00 |

[1] Wavelengths are given in Å.

[2] Optical Depth.

**Table 13. Predicted optical depth, sorted by wavelength for thermal-broadened model, for case II.**

| Crab Abund. | | Solar Abund. | | ISM Abund. | |
|---|---|---|---|---|---|
| Line[1] | Opt. Dpt.[2] | Line | Opt. Dpt. | Line | Opt. Dpt. |
| H I 1025 | 1.16E+00 | H I 1025 | 4.69E+00 | H I 1025 | 4.17E+00 |
| O I 1025 | 9.81E-01 | O I 1025 | 3.92E+00 | O I 1025 | 3.53E+00 |
| O VI 1031+1037 | 6.78E+01 | O VI 1031+1037 | 6.34E+01 | O VI 1031+1037 | 3.75E+01 |
| H I 1215 | 7.24E+00 | H I 1215 | 2.93E+01 | H I 1215 | 2.61E+01 |
| N V 1239+1243 | 3.04E+00 | N V 1239+1243 | 8.42E+00 | N V 1239+1243 | 7.06E+00 |
| C IV 1548+1551 | 5.50E+01 | C IV 1548+1551 | 8.61E+00 | C IV 1548+1551 | 7.79E+00 |

[1] Wavelengths are given in Å.

[2] Optical Depth.

**Table 14. Predicted optical depth, sorted by wavelength for thermal-broadened model, for case III.**

| Crab Abund. | | Solar Abund. | | ISM Abund. | |
|---|---|---|---|---|---|
| Line[1] | Opt. Dpt.[2] | Line | Opt. Dpt. | Line | Opt. Dpt. |
| H I 1025 | 3.51E+00 | H I 1025 | 1.31E+01 | O I 1025 | 9.75E+00 |
| O I 1025 | 2.86E+00 | O I 1025 | 1.06E+01 | H I 1025 | 1.18E+01 |
| O VI 1031+1037 | 1.34E+02 | O VI 1031+1037 | 1.03E+02 | O VI 1031+1037 | 6.24E+01 |
| H I 1215 | 2.19E+01 | H I 1215 | 8.17E+01 | H I 1215 | 7.37E+01 |
| N V 1239+1243 | 8.35E+00 | N V 1239+1243 | 1.99E+01 | N V 1239+1243 | 1.71E+01 |
| C IV 1548+1551 | 2.00E+02 | Si IV 1394 | 1.47E-01 | C IV 1548+1551 | 2.51E+01 |
| | | C IV 1548+1551 | 2.72E+01 | | |

[1] Wavelengths are given in Å.

[2]Optical Depth.

**Table 15. Line optical depths for static and dynamic cases (case I, solar)**

| Line | Thermal | $v$ =1680 km s$^{-1}$ |
|---|---|---|
| O VI 1031Å+1037Å | 38.1 | 1.81E-01 |
| C IV 1548Å+1551Å | 2.68 | 1.95E-02 |